\documentclass[graybox, envcountchap]{svmult}

\usepackage{mathptmx}        
\usepackage{amsmath}
\usepackage{color}
\usepackage{helvet}          
\usepackage{courier}         
\usepackage{dirtree}

\usepackage{makeidx}        
\usepackage{graphicx}        
\usepackage{subfig}

\usepackage{multicol}        
\usepackage[bottom]{footmisc}

\usepackage{hyperref}        
\hypersetup{colorlinks=true,urlcolor=blue}

\usepackage[misc]{ifsym}

\newcommand{\be}{\begin{eqnarray}}
\newcommand{\ee}{\end{eqnarray}}

\makeindex             

\begin{document}


\title{Search for Variations of Fundamental Constants}

\author{Cosimo Bambi}

\institute{Cosimo Bambi (\Letter) \at Center for Field Theory and Particle Physics and Department of Physics\\Fudan University, 200438 Shanghai, China\\ \email{bambi@fudan.edu.cn}}

\maketitle

\abstract{The possibility of variations of the values of fundamental constants is a phenomenon predicted by a number of scenarios beyond General Relativity. This can happen if ``our'' fundamental constants are not the actual constants of the fundamental theory and their value is instead determined, for example, by the background value of some new field or the size of extra dimensions. So far, most studies have been devoted to constrain possible temporal variations of the electromagnetic fine structure constant $\alpha$, the proton to electron mass ratio $\mu = m_p/m_e$, and the gravitational constant $G_{\rm N}$. Apart some claims of the detection of a temporal or spatial variation of $\alpha$ and $\mu$, so far there is no clear observational evidence of any variation of our fundamental constants.}


\section{Introduction}\label{sec:intro}

Our fundamental theories for the description of Nature are currently General Relativity and the Standard Model of particle physics. General Relativity describes the gravitational interactions and the spacetime structure and has 2~fundamental constants: the gravitational constant $G_{\rm N}$ and the cosmological constant $\Lambda$. The Standard Model of particle physics describes matter and the other known interactions (electromagnetic, strong, and weak interactions) and has 19~fundamental constants: 9~fermion masses, 3~quark mixing angles, 1~quark CP violating phase, 3~gauge couplings, 1~Higgs vacuum expectation value, 1~weak mixing angle, and 1~CP violating parameter of the strong interactions. Today we know that neutrinos have a non-vanishing mass and therefore we have to add at least 3~neutrino masses, 3~neutrino mixing angles, and 1~neutrino CP violating phase. Overall, we arrive at 28~fundamental constants. In principle, any other physical quantity (e.g. the proton and neutron masses) can be calculated from the theory and depends on these 28~fundamental constants.

However, both General Relativity and the Standard Model of particle physics may not be the true fundamental theories of Nature but rather some effective theories valid today at low energies in weak gravitational fields. In such a case, it is possible that the 28~fundamental constants listed above are not really fundamental constants but derived quantities that can be calculated from the theory, depend on the actual fundamental constants of the theory, and eventually may not be constant. From a purely aesthetic point of view, 28~fundamental constants sound too many, as one would like to have a theory with a very small number of inputs and derive any other physical quantity from the theory.

The search for variations of fundamental constants can be seen as a test of General Relativity because the variation of constants is a common phenomenon in a number of theories of gravity beyond General Relativity. For example, if there are extra dimensions, the fundamental theory and its fundamental constants must be defined in the entire spacetime. The $n$-dimensional Einstein-Hilbert action is
\be
S_n = \frac{M^{n-2}}{2} \int d^n x \, \sqrt{-g_n} \, R_n \, ,
\ee
where $M$ is the fundamental scale of the full theory and $g_n$ and $R_n$ are, respectively, the determinant of the $n$-dimensional metric and the $n$-dimensional Ricci scalar. After integrating over the extra dimensions, we should recover the standard 4-dimensional Einstein-Hilbert action
\be
S_4 = \frac{M_{\rm Pl}^2}{2} \int d^4 x \, \sqrt{-g_4} \, R_4 \, ,
\ee
where $M_{\rm Pl}$ is our Planck mass. The fundamental constant of the theory is $M$, not $M_{\rm Pl}$. The latter depends on $M$ and on the size and the properties of the extra dimensions. In a cosmological context, it is very natural to expect that even the extra dimensions evolve with time, and such a phenomenon should induce a variation of our 4-dimensional gravitational constant $G_{\rm N}$ over cosmological scales.

The variation of fundamental constants is a natural phenomenon in string theories~\cite{Maeda:1987ku,Damour:1994zq,Damour:1994ya}. The effective 4-dimensional Lagrangian of Yang-Mills fields normally looks like
\be
\mathcal{L}_{\rm YM} = - \frac{B(\phi)}{4} F^2 \, ,
\ee
where $B(\phi)$ is some function of the scalar field $\phi$. The effective 4-dimensional Yang-Mills coupling constant $g_{\rm YM}$ is thus  
\be
\frac{1}{g_{\rm YM}^2} = B(\phi) \, ,
\ee
 and it is constant only if $\phi$ is constant. However, one can expect that $\phi$ is not constant in an expanding universe or in different gravitational fields.

Studies on the variation of fundamental constants have mainly focused on the possibility of temporal variations. In part, this choice is motivated by the idea of the existence of some background field or extra dimensions that can naturally evolve in time with the expansion of the Universe. Actually, if such background field or extra dimensions really exist, there should be a mechanism to make them so stable. It would indeed be more natural to expect dramatic variations of fundamental constants that certainly we do not see. The possibility of spatial variations, either because the Universe is not isotropic or because the values of some fundamental constants can depend on the strength of the gravitational field, is less explored in the current literature.

Variations of fundamental constants can also lead to violations of the Weak Equivalence Principle, which asserts that the trajectory of any freely-falling test-body is independent of its internal structure and composition. The proton and neutron masses, as well as the masses of atoms, depend on the values of some fundamental constants in a non-trivial way. If our fundamental constants change in space and/or time, we can expect that the masses of bodies of different composition change in a different way, and this should lead to have freely-falling test-bodies with different trajectories.

If one of our fundamental constants can vary in space and/or time, it is quite natural to expect that even other fundamental constants can vary. This clearly depends on the underlying fundamental theory, which may be unknown. The exception may be the gravitational constant $G_{\rm N}$ since -- ignoring the cosmological constant $\Lambda$, whose effect in most physical systems is completely negligible -- it is the only constant of the gravity sector. If, for example, we have a variation of the fine structure constant $\alpha$, we can expect that at least all the gauge couplings of the Standard Model can vary. For this reason, if we want a clean measurement of the possible variation of a constant, we should study a physical process that depends only on the value of that fundamental constant and is independent of any other constant. Connections among different constants are natural but model-dependent. For example, if we believe in Grand Unification Theories, it is possible to link variations of $\alpha$ with variations of the gauge couplings of the Standard Model of particle physics~\cite{Calmet:2001nu}.

The aim of this chapter is to provide a short overview on the very vast topic of search for variations of fundamental constants. A complete review of this research field would require an entire book rather than a short chapter; longer reviews are Refs.~\cite{Uzan:2002vq,Uzan:2010pm,Martins:2017yxk}. There is a rich literature on the topic and these studies involve different branches of physics, including atomic physics, nuclear physics, particle physics, astrophysics, and gravity. In the past, there were some claims of possible evidence of variation of the fine structure constant $\alpha$ and of the proton to electron mass ratio $\mu = m_p/m_e$, which has certainly stimulated the interest of the scientific community on this topic. However, it is now thought that those non-vanishing measurements of variations of $\alpha$ and $\mu$ are likely related to systematic uncertainties not fully under control.


\section{Experimental and observational constraints}\label{sec:constraints}

\subsection{Atomic clocks}

Laboratory experiments can constrain the variations of fundamental constants by comparing two different atomic clocks (or an atomic clock with an ultra-stable oscillator). While laboratory experiments can study variations of fundamental constants on timescales of months to a few years, the precision of these measurements can be so good that their constraints can be very competitive.

In general, the characteristic frequency of an atomic clock depends somehow on the fine structure constant $\alpha$, the proton to electron mass ratio $\mu$, and the proton gyromagnetic factor $g_p$. If two atomic clocks have, respectively, characteristic frequency $\nu_1$ and $\nu_2$, their ratio can be written as
\be
\frac{\nu_1}{\nu_2} \propto \alpha^{\lambda_\alpha} \, \mu^{\lambda_\mu} \, g^{\lambda_g }
\ee
where $\lambda_\alpha$, $\lambda_\mu$, and $\lambda_g$ are given by
\be
\lambda_\alpha = \frac{d}{d \ln \alpha} \ln \left(\frac{\nu_1}{\nu_2}\right) \, , \quad
\lambda_\mu = \frac{d}{d \ln \mu} \ln \left(\frac{\nu_1}{\nu_2}\right) \, , \quad
\lambda_g = \frac{d}{d \ln g} \ln \left(\frac{\nu_1}{\nu_2}\right) \, . \quad
\ee
From the comparison of the two atomic clocks, we constrain a possible temporal variation of the ratio between the two frequencies, $\nu_1/\nu_2$. If we can calculate $\lambda_\alpha$, $\lambda_\mu$, and $\lambda_g$, we can connect the experimental constraint on $\nu_1/\nu_2$ to the constraint on a certain combination of $\alpha$, $\mu$, and $g_p$. If $\lambda_\mu=\lambda_g=0$, we can constrain the fine structure constant $\alpha$ without any assumption of possible temporal variations of $\mu$ and $g_p$.

Tab.~\ref{t-atomic} shows current laboratory constraints on possible variations of fundamental constants, where $\lambda_\alpha$, $\lambda_\mu$, and $\lambda_g$ are from Ref.~\cite{Martins:2017yxk}. The measurement reported in Rosenband et al. (2008)~\cite{Rosenband2008} constrain $\dot{\alpha}/\alpha$ because they have $\lambda_\mu=\lambda_g=0$. In the next years, laboratory experiments may be able to improve the constraints in Tab.~\ref{t-atomic} by several orders of magnitude.

\begin{table*}
\caption{Summary of current laboratory constraints on possible variations of fundamental constants. Shelkovnikov et al. (2008) compared the frequency of the fundamental hyperfine transition of $^{133}$Cs with the frequency of a rovibrational transition SF$_6$ from a CO$_2$ laser. Tamm et al. (2014) worked with the transition $^2$S$_{1/2}$$\rightarrow$$^2$D$_{3/2}$ of $^{171}$Yb$^+$, while Huntemann et al. (2014) used the transition $^2$S$_{1/2}$$\rightarrow$$^2$F$_{7/2}$ of $^{171}$Yb$^+$.}
{\renewcommand{\arraystretch}{1.3}
	\begin{tabular}{c c c c c c}
	    \hline
		\hline
Clocks & \hspace{0.1cm} $\frac{d}{dt} \ln \left(\frac{\nu_1}{\nu_2}\right)$ \hspace{0.1cm} & \hspace{0.3cm} $\lambda_\alpha$ \hspace{0.3cm} & \hspace{0.3cm} $\lambda_\mu$ \hspace{0.3cm} & \hspace{0.3cm} $\lambda_g$ \hspace{0.3cm} & Reference \\
\hline
$^{133}$Cs vs $^1$H & $(3.2 \pm 6.3) \cdot 10^{-15}$~yr$^{-1}$ & $2.83$ & $1$ & $-1.266$ & Fischer et al. (2004)~\cite{Fischer:2004jt} \\
$^{133}$Cs vs $^{199}$Hg$^+$ & $(-3.7 \pm 3.9) \cdot 10^{-16}$~yr$^{-1}$ & $5.77$ & $1$ & $-1.266$ & Fortier et al. (2007)~\cite{Fortier:2007jf} \\
$^{133}$Cs vs SF$_6$ & $(-1.9 \pm 2.7) \cdot 10^{-14}$~yr$^{-1}$ & $2.83$ & $0.5$ & $-1.266$ & Shelkovnikov et al. (2008)~\cite{Shelkovnikov:2008rv} \\
$^{199}$Hg$^+$ vs $^{27}$Al$^+$ & $(5.3 \pm 7.9) \cdot 10^{-17}$~yr$^{-1}$ & $-2.95$ & $0$ & $0$ & Rosenband et al. (2008)~\cite{Rosenband2008}\\
$^{133}$Cs vs $^{87}$Rb & $(1.07 \pm 0.49) \cdot 10^{-16}$~yr$^{-1}$ & $0.49$ & $0$ & $-2$ & Gu\'ena et al. (2012)~\cite{Guena:2012zz} \\
$^{162}$Dy vs $^{164}$Dy & $(-5.8 \pm 6.9) \cdot 10^{-17}$~yr$^{-1}$ & $1$ & $0$ & $0$ & Leefer et al. (2013)~\cite{Leefer:2013waa} \\
$^{133}$Cs vs $^{171}$Yb$^+$ & $(-0.5 \pm 1.9) \cdot 10^{-16}$~yr$^{-1}$ & $1.83$ & $1$ & $-1.266$ & Tamm et al. (2014)~\cite{Tamm2014} \\
$^{133}$Cs vs $^{171}$Yb$^+$ & $(-0.2 \pm 4.1) \cdot 10^{-16}$~yr$^{-1}$ & $8.83$ & $1$ & $-1.266$ & Huntemann et al. (2014)~\cite{Huntemann:2014dya} \\
\hline
\hline
	\end{tabular}}
	\label{t-atomic}
\end{table*}


\subsection{The Oklo natural nuclear reactor}

The Oklo uranium mine is situated in Gabon and, thanks to its peculiar conditions, about 2~Gyr ago (corresponding to a redshift $z \approx 0.14$) it acted as a natural nuclear reactor for a few million years~\cite{Naudet}. Alexander Shlyakhter was the first to point out that the study of the Oklo nuclear reactor could test the temporal variation of fundamental constants~\cite{Shlyakhter:1976}.

The isotopes of samarium $^{147}$Sm and $^{149}$Sm are not fission products. In Oklo ores, the ratio $^{149}$Sm/$^{147}$Sm is about 0.02, while it is about 0.9 in normal samarium. Such a low value of $^{149}$Sm/$^{147}$Sm is thought to result from the following neutron capture process
\be\label{eq-sm}
n + {}^{149}{\rm Sm} \rightarrow {}^{150}{\rm Sm} + \gamma \, ,
\ee
where the flux of neutrons is produced by fission processes in the Oklo nuclear reactor. The key point is that the cross-section of this process is dominated by a capture resonance resulting from an almost perfect cancellation between the electromagnetic and strong interactions. The cross-section of this reaction is approximated well by the Breit-Wigner formula
\be\label{eq-sigma}
\sigma_n (E) = \frac{g \pi}{2} \frac{\hbar^2}{m_n E} \frac{\Gamma_n \Gamma_\gamma}{\left( E - E_R\right)^2 + \Gamma_{\rm tot}^2/4}
\ee
where the statistical factor is $g = 9/16$ for the reaction in~(\ref{eq-sm}). The total width is $\Gamma_{\rm tot} = \Gamma_n + \Gamma_\gamma$, and $\Gamma_n$ and $\Gamma_\gamma$ are, respectively, the neutron and the radiative partial widths. $E_R$ is the resonance energy and for the reaction in~(\ref{eq-sm}) its value today is 97.3~meV.

The cross-section $\sigma_n$ in Eq.~(\ref{eq-sigma}) cannot be measured directly. We can define an effective cross-section, $\hat{\sigma}_n$, which includes the geometry of the reactor zone and the properties of the neutron flux. The evolution of the number density of the isotopes in the Oklo nuclear reactor is governed by a system of differential equations in which we have $\hat{\sigma}_n$. Assuming that the cross-sections of the reactions and the neutron flux were constant during the period in which the reactor was active, we can solve the system and find the value of $\hat{\sigma}_n$ at the time of the nuclear reactor. To infer $\sigma_n$ and then the resonance energy $E_R$, it is necessary to specify the geometry of the reactor zone and the properties of the neutron flux.

Once we have the value of the resonance energy $E_R$ at the time of the Oklo nuclear reactor, we need a model to connect the value of the resonance energy to the values of the fundamental constants. In Ref.~\cite{Damour:1996zw}, the authors consider that only the fine structure constant $\alpha$ could vary and in their analysis find the following 2~$\sigma$ constraint
\be
- 0.9 \cdot 10^{-7} < \frac{\Delta\alpha}{\alpha} < 1.2 \cdot 10^{-7} \, ,
\ee
where $\Delta\alpha$ is the variation of $\alpha$ at the time of the Oklo nuclear reactor with respect to its value today. For $\dot{\alpha} \equiv d\alpha/dt = {\rm constant}$, we have
\be
- 6.7 \cdot 10^{-17}~{\rm yr}^{-1} < \frac{\dot{\alpha}}{\alpha} < 5.0 \cdot 10^{-17}~{\rm yr}^{-1} \, .
\ee

In Ref.~\cite{Fujii:1998kn}, the authors obtain two solutions for the variation of the resonance energy $\Delta E_R$: a solution consistent with no variation $\Delta E_R = 9 \pm 11$~meV and another solution consistent with a non-vanishing variation $\Delta E_R = -97 \pm 8$~meV. The first solution consistent with no variation gives the following constraints on the fine structure constant $\alpha$
\be
\frac{\Delta\alpha}{\alpha} &=& \left( -0.36 \pm 1.44 \right) \cdot 10^{-8} \, , \\
\frac{\dot{\alpha}}{\alpha} &=& \left( -0.2 \pm 0.8 \right) \cdot 10^{-17}~{\rm yr}^{-1} \, .
\ee

The constraints above hold in the case only the value of the fine structure constant $\alpha$ can potentially be different from its value today, while all other fundamental constants are supposed to have the same value as today. In Ref.~\cite{Olive:2002tz}, the authors consider a scenario inspired by Grant Unification Theories in which the values of all gauge and Yukawa couplings of the Standard Model of particle physics would be connected and a variation of $\alpha$ implies a variation of all other constants. In this way, a small variation of $\alpha$ is amplified by the physics of strong interactions and eventually one can get a constraint on $\Delta\alpha/\alpha$ about two orders of magnitude more stringent.


\subsection{Quasar absorption spectra}

The search for temporal variations of fundamental constants in absorption spectra of high-redshift quasars is probably one of the most explored methods to find variations of fundamental constants. This technique can provide very competitive constraints and, at the same time, in the past years there were a few studies claiming the measurement of a temporal variation of the fine structure constant $\alpha$ and the proton to electron mass ratio $\mu$. This kind of measurements are certainly challenging, as there are many potential sources of systematic uncertainties that can affect the final result~\cite{Murphy:2000nr}.

Within the non-relativistic approximation, all atomic transitions have the same dependence of the fine structure constant $\alpha$, so it is impossible to distinguish an energy shift caused by a variation of $\alpha$ from conventional Doppler, cosmological, and gravitational shifts. However, relativistic effects can break such a degeneracy. For example, fine structure doublets have the following frequency splitting
\be
\Delta \nu = \frac{\alpha^2 Z^4 R_\infty}{2 n^3}~{\rm cm}^{-1} \, ,
\ee
where $Z$ is the atomic number of the atom, $R_\infty$ is the Rydberg constant, and $n$ is the principal quantum number. In this case, one has to compare the wavelength splitting in the astrophysical source with the laboratory measurement. Malcolm Savedoff was the first to use the fine structure splitting to constrain a possible variation of the fine structure constant $\alpha$~\cite{Savedoff1956}. He studied the spectrum of the radio source Cygnus~A at redshift $z \approx 0.057$ and found
\be
\frac{\Delta \alpha}{\alpha} = \left( 1.8 \pm 1.6 \right) \cdot 10^{-3} \, .
\ee

A more sophisticated method to search for possible temporal variations of $\alpha$ is the combined analysis of the absorption lines of a set of multiplets of different ions~\cite{Dzuba:1998au,Dzuba:1999zz,Webb:1998cq}. With this method, the authors of Ref.~\cite{Webb:1998cq} considered a sample of 30~absorption systems with redshift $0.5 < z < 1.6$ finding
\be
\frac{\Delta \alpha}{\alpha} &=& \left( -0.2 \pm 0.4 \right) \cdot 10^{-5} \quad {\rm for} \,\,\, 0.5 < z < 1 \, , \\
\frac{\Delta \alpha}{\alpha} &=& \left( -1.9 \pm 0.5 \right) \cdot 10^{-5} \quad {\rm for} \,\,\, 1 < z < 1.6 \, .
\ee
For $z > 1$, the detection of a non-vanishing variation of the fine structure constant is at more than 3~$\sigma$. Such a result was confirmed by the same group in other studies~\cite{Murphy:2000pz,Webb:2000mn}, but was not confirmed by the authors of Ref.~\cite{Srianand:2004mq}, who considered a smaller sample of high-quality quasar spectra and found
\be
\frac{\Delta \alpha}{\alpha} = \left( -0.06 \pm 0.06 \right) \cdot 10^{-5} \, .
\ee
Comments on the results of these two groups were reported in Refs.~\cite{Murphy:2007qs,Srianand:2007zz}. These two independent measurements were obtained from two different telescopes (Keck/HIRES vs VLT/UVES) observing different parts of the sky (Northern vs Southern hemispheres). In Ref.~\cite{Webb:2010hc}, the authors of Ref.~\cite{Webb:1998cq} argued that the combined analysis of Keck/HIRES and VLT/UVES data suggests a spatial variation of the fine structure constant $\alpha$ at more than 4~$\sigma$.

Constraints on the temporal evolution of the proton to electron mass ratio $\mu$ can be inferred from the study of molecular lines, since rotational transitions are proportional to $1/\mu$ and vibrational transition to $1/\sqrt{\mu}$~\cite{Thompson1975}. From the combined analysis of two absorption system at redshift $z=2.5947$ and $z=3.0249$, the authors of Ref.~\cite{Ivanchik:2005ws} reported two measurements of a non-vanishing variation of $\mu$
\be
\frac{\Delta \mu}{\mu} = \left( 1.7 \pm 0.7 \right) \cdot 10^{-5} \, , \quad
\frac{\Delta \mu}{\mu} = \left( 3.1 \pm 0.8 \right) \cdot 10^{-5} \, ,
\ee
where the two measurements refer to two different laboratory measurements of the transitions. Some studies confirmed such a non-null measurement but other studies did not. For example, in Ref.~\cite{Reinhold:2006zn} the authors reported the measurement
\be
\frac{\Delta \mu}{\mu} = \left( 2.4 \pm 0.6 \right) \cdot 10^{-5} \, .
\ee
These measurements are challenging because of the numerous sources of systematic uncertainties, which are difficult to reduce and properly take into account.

To conclude this subsection, there are a few claims in the literature of the measurement of a non-vanishing variation of the fine structure constant $\alpha$ and the proton to electron mass ratio $\mu$ from quasar absorption spectra. The common consensus is that these non-null measurements are due to systematic uncertainties not completely under control, and work is on-going to improve these measurements and understand the systematics behind.


\subsection{Cosmic microwave background}

After the big bang nucleosynthesis, the primordial plasma was mainly made of photons, protons, helium-4 nuclei, and electrons. The abundances of other light elements were quite low. Neutrinos and other possible particles beyond the Standard Model did not interact with the primordial plasma. The process
\be
p + e^- \leftrightarrow H + \gamma 
\ee
was in equilibrium, but it was impossible to create neutral hydrogen because newly formed hydrogen atoms were immediately destroyed by photons. The event in which protons and electrons form neutral hydrogen is called the recombination and, in the standard framework, it happens when the temperature of the Universe is $T_{\rm rec} \approx 0.3$~eV, which is much lower than then hydrogen ionization energy $E_{\rm ion} = 13.6$~eV because of the large number of photons with respect to protons and electrons. Before recombination, photons and matter were in thermal equilibrium through elastic Thomson scattering of photons off free electrons. The Thomson cross-section is
\be
\sigma_{\rm} = \frac{8\pi}{3} \frac{\alpha^2 \hbar^2}{m_e^2 c^2} \, ,
\ee
where $m_e$ is the electron mass. After recombination, photons decouple from matter because the cross-section of photon elastic scattering on a neutral atom is much smaller than the Thomson cross-section and the expansion rate of the Universe exceeded the scattering rate: this event is called the decoupling and approximately occurred at the time of recombination (but both events were not instantaneous).

The values of the fine structure constant $\alpha$ and of the electron mass $m_e$ determine the temperature of recombination and, in turn, the temperature of decoupling, as well as the residual ionization after decoupling. Both effects have an impact on the temperature and polarization anisotropies of the cosmic microwave background. There are a few studies published in the literature~\cite{Hannestad:1998xp,Kaplinghat:1998ry,Avelino:2000ea}. Even the gravitational constant $G_{\rm N}$ plays an important role~\cite{Riazuelo:2001mg}. From the Plank 2018 data, the authors of Ref.~\cite{Hart:2019dxi} find
\be
\frac{\Delta\alpha}{\alpha} &=& 0.0005 \pm 0.0024 \, , \\
\frac{\Delta m_e}{m_e} &=& 0.112 \pm 0.059 \, ,
\ee
assuming, respectively, a possible variation of $\alpha$ with constant $m_e$ and a possible variation of $m_e$ with constant $\alpha$. Adding baryon acoustic oscillation data, the measurement of the electron mass becomes $\Delta m_e/m_e = 0.0078 \pm 0.0067$. Including also the supernova data, they find $\Delta m_e/m_e = 0.0190 \pm 0.0055$.


\subsection{Big bang nucleosynthesis}

The big bang nucleosynthesis is the production of light elements (mainly $^4$He, D, $^3$He, and $^7$Li) in the very early Universe; see, for instance, Ref.~\cite{Bambi:2015mba}. It is the result of the competition between a large number of particle and nuclear processes and the expansion rate of the Universe, which dilutes and cools down the primordial plasma. Theoretical calculations can predict the primordial abundances of $^4$He, D, $^3$He, and $^7$Li, which can then be compared with the estimates of the primordial abundances from observational data. This is not straightforward because the data are subject to poorly known systematic errors and evolutionary effects~\cite{Steigman:2005wb}. As of now, the big bang nucleosynthesis is the event that permits us to go further back in time and test the physics when the Universe was only 1~second old.

The calculations of the primordial abundance of $^4$He can be sketched as follows. At high temperatures ($T > 1$~MeV), protons and neutrons are in statistical equilibrium with the primordial plasma through the following weak processes
\be\label{eq-weak}
p + \bar{\nu}_e \leftrightarrow n + e^+ \, , \quad
p + e^- \leftrightarrow n + \nu_e \, , \quad
n \rightarrow p + e^- + \bar{\nu}_e \, ,
\ee
and the ratio of the neutron to proton number density is $n_n / n_p \approx e^{\Delta m / T}$, where $\Delta m = m_n - m_p = 1.29$~MeV is the difference between the neutron and proton masses. Protons and neutrons remain in equilibrium as long as the rate of the weak interactions in Eq.~(\ref{eq-weak}), $\Gamma_{\rm weak}$, exceeds the expansion rate of the Universe, $H(T)$, where $H(T)$ is the Hubble parameter at the temperature $T$
\be\label{eq-hubble}
H(T) = \left( \frac{4 \pi^3 G_{\rm N}}{45} g_* \right)^{1/2} T^2 \, ,
\ee
and $g_*$ is the effective number of relativistic degrees of freedom in the plasma. When $\Gamma_{\rm weak} < H(T)$, we have the freeze-out of the weak interactions: the value of $n_n / n_p$ remains constant except for the fact that neutrons decay into protons. In the framework of standard physics, the temperature of freeze-out of the weak interactions is around 1~MeV.

The first step to produce light elements is the synthesis of deuterium. At high temperatures, the process
\be
p + n \leftrightarrow {\rm D} + \gamma \, 
\ee
cannot produce any significant amount of deuterium because the deuterium is quickly destroyed by photons. The deuterium binding energy is $W = m_p + m_n - m_D = 2.225$~MeV, but since the number density of photons in the plasma is much higher than the number density of $p$, $n$, and D, it is necessary that the temperature of the Universe is significantly lower than $W$ before the synthesis of deuterium becomes efficient. In the framework of standard physics, the temperature of the deuterium synthesis is around 70~keV. At first approximation, the primordial abundance of $^4$He can be estimated assuming that all neutrons survived at the deuterium synthesis contribute to create $^4$He. Numerical calculations are required for more precise estimates of the primordial abundance of $^4$He as well as for the estimates of the primordial abundances of D, $^3$He, and $^7$Li.

The values of the fundamental constants at the time of the big bang nucleosynthesis can have a dramatic impact on the primordial abundances of $^4$He, D, $^3$He, and $^7$Li. A temporal variation of the value of the gravitational constant $G_{\rm N}$ affects the expansion rate of the Universe. However, a simple modification of the value of $G_{\rm N}$ in Eq.~(\ref{eq-hubble}) is not self-consistent because that expression is obtained within General Relativity where $G_{\rm N}$ is constant. In models predicting an effective variation of the gravitational constant, one can expect that the expansion rate of the Universe includes extra terms with the temporal derivatives of $G_{\rm N}$. Neglecting those terms corresponds to assume that the variation of $G_{\rm N}$ is slow during the big bang nucleosynthesis and that we can approximate the expansion rate of the Universe with Eq.~(\ref{eq-hubble}) with a different value for $G_{\rm N}$. There are a number of publications in the literature reporting studies of the impact of the value of $G_{\rm N}$ during the big bang nucleosynthesis and its observational constraints. Some studies consider specific theoretical models with the corresponding modified Hubble parameter and other studies follow a more phenomenological approach in which the theory is not specified~\cite{Accetta:1990au,Damour:1998ae,Copi:2003xd,Bambi:2005fi,Clifton:2005xr}. In general, the primordial abundance of $^4$He provides the most reliable constraints: a larger value of $G_{\rm N}$ increases the primordial abundance of $^4$He and a lower value of $G_{\rm N}$ has the opposite effect. Depending on the specific model and measurement of primordial abundance of $^4$He, we can typically conclude that the value of $G_{\rm N}$ at the big bang nucleosynthesis could not differ from the current one more than $\sim$10\%.

The big bang nucleosynthesis can be used, and has been used, even to constrain the variations of other fundamental constants~\cite{Kolb:1985sj,Campbell:1994bf,Bergstrom:1999wm,Ichikawa:2002bt,Coc:2012xk,Clara:2020efx}. The strong point of the big bang nucleosynthesis is that we can test the physics when the Universe was only 1~second old and currently we do not have any other robust method to go to even earlier times. The weak point is that there are so many processes involving all known interactions (electromagnetic, strong, and weak interactions, as well as gravity) that it is impossible to derive a clean and model-independent constraint. For example, if we consider the possibility of a variation of the fine structure constant $\alpha$, we can likely expect that many other fundamental constants change value, but in this case we need some relationship among them. Moreover, the value of $\alpha$ affects many physical quantities, but many inputs in the standard big bang nucleosynthesis are obtained from laboratory measurements and we do not have a so good knowledge of nuclear physics to derive the values of these quantities from the theory. Even the large number of cross-sections required in the calculations of the primordial abundances are measured in laboratory and it is not straightforward to derive their values in the case of variations of fundamental constants.


\subsection{Compact objects}

In strong gravitational fields, like those generated around compact objects, the values of fundamental constants may be different from those we can measure in our laboratories on Earth; see, e.g., Ref.~\cite{Davis:2016avf}.

In Refs.~\cite{Berengut:2013dta,Hu:2020zeq}, the authors compare laboratory spectra with white dwarf spectra to estimate possible differences between the values of the fine structure constant $\alpha$ in the strong gravitational field of the surface of a white dwarf and in our laboratories on Earth. The difference between the dimensionless gravitational potential on a white dwarf and on Earth is $\Delta\phi \approx 5 \cdot 10^{-5}$. From two independent sets of laboratory Fe~V wavelengths, they find two measurements~\cite{Hu:2020zeq}
\be
\frac{\Delta\alpha}{\alpha} = \left(6.4 \pm 2.3 \right) \cdot 10^{-5} \, , \qquad
\frac{\Delta\alpha}{\alpha} = \left(4.2 \pm 2.8 \right) \cdot 10^{-5} \, ,
\ee
where the uncertainties here are supposed to include both statistical and systematic errors. The total uncertainties are dominated by the systematic ones, and the latter are dominated by the precision of laboratory wavelength measurements. These measurements differ from a null result by 2.8~$\sigma$ and 1.5~$\sigma$, respectively~\cite{Hu:2020zeq}.

Stronger gravitational fields can be found around black holes. In Ref.~\cite{Hees:2020gda}, the authors study the spectra of 5~S-stars orbiting the supermassive black hole at the center of our Galaxy to measure possible variations of the fine structure constant $\alpha$. However, their stars are relatively far from the central object and, in the end, one can probe a gravitational field as strong as on the surface of a white dwarf. For example, from the star S0-6 they find
\be
\frac{\Delta\alpha}{\alpha} = \left(1.0 \pm 1.2 \right) \cdot 10^{-4} \quad {\rm for} \,\,\,\, \phi = 2.4 \cdot 10^{-6} \, .
\ee

The analysis of fluorescent emission lines in the reflection spectra of black holes can potentially test variations of fundamental constants in gravitational fields with gravitational potentials up to $\phi \sim 1$, as suggested in Ref.~\cite{Bambi:2013mha}. X-ray reflection spectroscopy~\cite{Bambi:2020jpe} has been already extensively used to test new physics with accreting black holes; see, e.g., Refs.~\cite{Bambi:2016sac,Cao:2017kdq,Tripathi:2020yts} and Chapter~5 in this book~\cite{Bambi:2022dtw}. However, the construction of a theoretical model to analyze a full reflection spectrum in the presence of variations of fundamental constants like $\alpha$ is not straightforward and currently there are no constraints reported in the literature with this method.

Compact objects have also been used to constrain temporal variations of fundamental constants. In a binary system, a temporal variation of the gravitational constant $G_{\rm N}$ would induce a variation of the orbital period. For a compact object, a non-vanishing $\dot{G}_{\rm N}$ leads also to a non-negligible temporal variation of the gravitational self-force, which is theory-depedent~\cite{Eardley1975,Nordtvedt:1990zz,Damour:1988zz,Kaspi:1994hp}. From the double pulsar PSR~J0737--3039A/B, today we have the constraint~\cite{Kramer:2021jcw}
\be
\frac{\dot{G}_{\rm N}}{G_{\rm N}} < \left(-0.8 \pm 1.4 \right) \cdot 10^{-13} \frac{1}{\mathcal{F}_{AB}}~{\rm yr}^{-1} \, , 
\ee
where $\mathcal{F}_{AB}$ is related to the corrections for the gravitational self-force (and $\mathcal{F}_{AB} \approx 1$ for non-compact objects like planets and stars)~\cite{Nordtvedt:1990zz,Damour:1988zz}.

In Ref.~\cite{Thorsett:1996fr}, the author argues that the average neutron star mass should be close to the Chandrasekhar limit at the time of formation of the compact object. The Chandrasekhar limit is technically the maximum mass for a white dwarf and reads
\be
M_{\rm Ch} \sim \frac{1}{m_N^2} \left(\frac{\hbar \, c}{G_{\rm N}}\right)^{3/2} \, ,
\ee
where $m_N$ is the nucleon mass. If the gravitational constant $G_{\rm N}$ was different in the past, this would have been recorded in old neutron stars, as $\dot{G}_{\rm N}/G_{\rm N} = - 2\dot{M}_{\rm Ch}/3M_{\rm Ch}$. Precise measurements of neutron star masses can be obtained from double neutron star binary pulsars. An upper limit to the age of a pulsar can be inferred from its spin down rate, and it turns out that most objects are young. However, for PSR~J1518+4904 one finds $\sim$16~Gyr. Assuming that the age of PSR~J1518+4904 is 10~Gyr (corresponding to the age of the Galactic disk) and noting that the masses of its two neutron stars are quite similar to that of young neutron stars, one can find the constraint (at 95\% CL)~\cite{Thorsett:1996fr}
\be
\frac{\dot{G}_{\rm N}}{G_{\rm N}} = \left( -0.6 \pm 4.2 \right) \cdot 10^{-12}~{\rm yr}^{-1} \, .
\ee


\begin{acknowledgement}
This work was supported by the National Natural Science Foundation of China (NSFC), Grant No.~12250610185 and Grant No. 11973019, the Natural Science Foundation of Shanghai, Grant No. 22ZR1403400, the Shanghai Municipal Education Commission, Grant No. 2019-01-07-00-07-E00035, and Fudan University, Grant No. JIH1512604.
\end{acknowledgement}



\end{document}